\documentclass[12pt]{article}
\usepackage{esfconf}
\usepackage{epsfig}

\begin{document}

\title{Influence of a medium on coherent scattering of high-energy photon}

\authors{V.N.Baier\adref{1}}
\addresses{\1ad Budker Institute of Nuclear Physics, Novosibirsk, 630090}

\maketitle

\begin{abstract}
The coherent scattering of photon in the Coulomb field
(the Delbr\"uck scattering) is considered for the momentum
transfer $\Delta \ll m$
in the frame of the quasiclassical operator method.
In high-energy region this process occurs over rather long distance.
The process amplitude is calculated taking into account the
multiple scattering of particles of the intermediate
electron-positron pair in a medium.
The result is the suppression of the process.

\end{abstract}
\vspace{0.2cm}

We with E.S. were acquainted during many decades. Conversations and 
discussions with him of scientific and non-scientific problems always were 
instructive and brought satisfaction and pleasure. E.S. was always absorbed
into physics and served as an example for more young physicists. The
tremendous success of this Memorial Conference shows how great 
the legacy of E.S.Fradkin.

1. The nonlinear effects of QED are due to the interaction of a photon with
electron-positron field. These processes are the photon-photon
scattering,  the coherent photon scattering, 
the photon splitting into two photons, and
the coalescence of two photons into photon in the Coulomb field.

History of the coherent photon scattering study can be found in 
review \cite{d3}. There is a special interest to 
the process for heavy elements
because contributions of higher orders of $Z\alpha$ ($Z|e|$ is the charge
of nucleus, $e^2=\alpha=1/137,~\hbar=c=1 $) into the amplitude of
photon scattering are very important. This means that one needs the theory
which is exact with respect to the parameter $Z\alpha$.
The amplitudes of the coherent photon scattering valid for any $Z\alpha$
for high photon energy $\omega \gg m$ and small scattering angle
(or small momentum transfer $\Delta$) were calculated by Cheng and Wu 
\cite{CW}. The approximate method of summing of the set of Feynman diagrams
with an arbitrary number of photons exchanged with the Coulomb source was
used. Another representation of these amplitudes 
was found by Milstein and Strakhovenko \cite{MS1},
using the quasiclassical Green function of the Dirac equation
in the Coulomb field. The Green function in a spherically symmetrical
external field was obtained by Lee and Milstein \cite{LM1}
where the coherent photon scattering in the screened Coulomb potential
was investigated as well. Lately this Green function was calculated for
"localized potential", and the coherent photon scattering was analyzed
using it by Lee, Milstein and Strakhovenko \cite{LM2}. 
Recently the process of the coherent photon scattering
was considered by Katkov and Strakhovenko \cite{KS} in frame 
of the quasiclassical operator method (see e.g.
\cite{BKS}) which appears to be very adequate for consideration of this
problem. This note is based on the recent paper of Katkov and author \cite{BK}.

The coherent photon scattering belong to the class of
electromagnetic processes
which in high-energy region occurs over rather long distance,
known as the formation length. Among other processes there are
the bremsstrahlung and the pair creation by a photon.
If anything happens to an electron
while travelling this distance, the emission can be disrupted.
Landau and Pomeranchuk~\cite{LP} showed that if the formation
length of bremsstrahlung becomes comparable to the distance over which
a mean angle of the multiple scattering becomes comparable with a
characteristic angle of radiation, the bremsstrahlung will be
suppressed. Migdal~\cite{M1} developed a quantitative
theory of this phenomenon which is known as the
Landau-Pomeranchuk-Migdal (LPM) effect.

Recently Katkov and  author \cite{L1, L3, L4} developed the new
approach to the theory of the LPM effect
in frame of the quasiclassical operator method.
In it the cross section of bremsstrahlung process
in the photon energies region where the influence of
the LPM effect is very strong
was calculated with term $\propto 1/L$ , where $L$
is characteristic logarithm of the problem,
and with the Coulomb corrections
taken into account. In the photon energy region, where the LPM effect
is "turned off", the obtained cross section
gives the exact Bethe-Maximon cross section (within power accuracy) with
the Coulomb corrections.  
This important feature was absent in the previous calculations.
We have analyzed the soft part of the spectrum, including all the 
accompanying effects. Perfect agreement
of the theory and SLAC data (see review \cite{E2}) 
was achieved in the whole interval of measured
photon energies.
Recently we apply this approach to the process
of pair creation by a photon \cite{L2}.

2. In the quasiclassical approximation the amplitude $M$
of the coherent photon scattering is described by diagram where
the electron-positron pair is created by the initial photon
with 4-momentum $k_1~(\omega,{\bf k}_1)$ and then annihilate into the final
photon with 4-momentum $k_2~(\omega,{\bf k}_2)$,~ 
so that the photon momentum transfer is $\Delta=|{\bf k}_2-{\bf k}_1|$. 
For high energy photon $\omega \gg m$
this process occurs over a rather long distance. 

It is convenient to describe the process of photon scattering in terms
of helicity amplitudes.There are two independent helicity amplitudes:
\[
M_{++}=M_{--},\quad M_{+-}=M_{-+},
\]
where the first subscript is the helicity of the initial photon and
the second is the helicity of the final photon.
When the initial photons are unpolarized the differential cross section
of scattering summed over final photons polarisation contains the
combination
\begin{equation}
2[|M_{++}|^2+|M_{+-}|^2].
\label{3.34}
\end{equation}
The imaginary part of helicity amplitudes can be written in the form
\cite{BK}
\begin{equation}
{\rm Im}~M_{\lambda \lambda'}=\frac{4Z^2\alpha^3 \omega}{m^2}
\int_{0}^{1}dx\int_{0}^{1}dy~ \mu_{\lambda \lambda'}f_{\lambda \lambda'},
\label{3.46}
\end{equation}
where
\begin{eqnarray}
&& \mu_{++}=1-2x(1-x)+4x(1-x)y(1-y),\quad \mu_{+-}=x(1-x)y^2;
\nonumber \\
&& f_{++}=\ln (ma)-\frac{2s^2+1}{2s\sqrt{1+s^2}}
\ln \left(s+\sqrt{1+s^2} \right)
-f(Z\alpha)+\frac{41}{42},\quad s=\frac{\Delta a}{2},
\nonumber \\
&&f_{+-}=1-\frac{1}{s\sqrt{1+s^2}}
\ln \left (s+\sqrt{1+s^2}\right),~
f(\xi)=\xi^2\sum_{n=1}^{\infty}\frac{1}{n(n^2+\xi^2)},
\label{3.47}
\end{eqnarray}
where $a$ is the screening radius of atom.
The important property of Eq.(\ref{3.46}) is that the dependence
on the screening radius $a$ originates in it from the Born approximation.
In this approximation in the case of arbitrary screening
the radius $a$ enters only in the combination
\begin{equation}
\frac{1}{a^2}+q_{\parallel}^2,\quad q_{\parallel}
=\frac{q_{m}}{x(1-x)y},
\quad q_m=\frac{m^2}{2\omega}.
\label{3.48}
\end{equation}
Because of this we can extend Eq.(\ref{3.47}) on the
case of arbitrary screening making the substitution
\begin{equation}
\frac{1}{a} \rightarrow \sqrt{q_{\parallel}^2+a^{-2}} \equiv
q_{ef},\quad s=\frac{\Delta}{2q_{ef}}
\label{3.49}
\end{equation}

In the case $a \gg \omega/m^2$ (the screening radius is very large, or
in other words we consider the photon scattering in the Coulomb field)
we have to substitute in (\ref{3.47})
\begin{equation}
a \rightarrow \frac{1}{q_{\parallel}},\quad s \rightarrow
 s_c=\frac{\Delta}{2q_{\parallel}}
=\frac{\Delta \omega}{m^2}x(1-x)y.
\label{3.56}
\end{equation}

In the case of the complete screening ($a \ll \omega/m^2$)
the functions
$f_{\lambda \lambda'}$ are independent of $x$ and $y$ and the corresponding
integrals are
\begin{equation}
\int_{0}^{1}dx\int_{0}^{1}dy~\mu_{++}=\frac{7}{9},\quad
\int_{0}^{1}dx\int_{0}^{1}dy~\mu_{+-}=\frac{1}{18}.
\label{3.54}
\end{equation}
For the scattering amplitudes we have
\begin{eqnarray}
&& {\rm Im}~M_{++}=\frac{28Z^2\alpha^3\omega}{9m^2}f_{++},\quad
{\rm Im}~M_{+-}=\frac{2Z^2\alpha^3\omega}{9m^2}f_{+-},
\nonumber \\
&&{\rm Re}~M_{++}=0,\quad {\rm Re}~M_{+-}=0.
\label{3.55}
\end{eqnarray}
The real part of amplitudes are calculated using dispersion relations.
The photon scattering amplitude in this case for arbitrary value of
parameter $am^2/2\omega$ was found recently in \cite{LM2}.

3. When a photon is propagating in a medium it
dissociates with probability $\propto \alpha$ into an electron-positron
pair. The virtual electron and positron interact with a medium
and can scatter on atoms. In this scattering 
the electron and positron interaction with the Coulomb field
in the course of the coherent scattering of photon is involved also.
There is a direct analogue with the LPM
effect: the influence of the multiple scattering on process of the
bremsstrahlung and pair creation by a photon in a medium at high energy. 
However there is the difference: in the LPM effect
the particles of electron-positron pair created by a
photon are on the mass shell while in the process
of the coherent scattering of photon this particles are off the mass shell,
but in the high energy region (this is the only region where the influence
of the multiple scattering is pronounced) the shift from the mass shell
is relatively small.
To include this scattering into consideration the amplitude
of the coherent scattering of photon
should be averaged over all possible trajectories of electron and positron.
This operation can be performed with the aid of the distribution function
averaged over the atomic positions of scatterer in the medium.

The photon scattering amplitudes can be written in the form
\begin{equation}
M_{++}=M_{++}^c+M_{++}^{(1)},\quad M_{+-}=M_{+-}^c+M_{+-}^{(1)},
\label{4.16}
\end{equation}
where $M_{++}^c$ is the main (logarithmic) term
and $M_{++}^{(1)}$ is the first correction to the scattering amplitude.
The main terms are \cite{BK}
\begin{eqnarray}
\hspace{-13mm}&& M_{++}^c=\frac{\alpha m^2\omega}{2\pi n_a} \int_{}^{}
\frac{d\varepsilon}{\varepsilon \varepsilon'}
\Phi_s(\nu_s),~ M_{+-}^c=0,~\nu_s=2\sqrt{iq_s},~q_s=QL_s,
\nonumber \\
\hspace{-13mm}&&\displaystyle{\Phi_s(\nu_s)=
=s_1\left(\ln p-\psi\left(p+\frac{1}{2}\right) \right)
+s_2\left(\psi (p) -\ln p+\frac{1}{2p}\right)},~p=i/(2\nu_s),
\nonumber \\
\hspace{-13mm}&&~Q=\frac{2\pi Z^2\alpha^2\varepsilon \varepsilon'n_a}
{m^4\omega}, 
~L_s(\varrho_c)=\ln \frac{a_{s2}^2}{\lambda_c^2 \varrho_c^2}-
F_2\left(\frac{\beta}{2} \right),~
\frac{a_{s2}}{\lambda_c}=183Z^{-1/3}{\rm e}^{-f},
\nonumber \\
\hspace{-13mm}&&F_2(z)=\frac{2z^2+1}{z\sqrt{1+z^2}}
\ln \left(z+\sqrt{1+z^2} \right)-1,~
s_1=1,~s_2=\frac{\varepsilon^2+\varepsilon'^2}{\omega^2},~\beta=a \Delta,
\label{4.17}
\end{eqnarray}
where $\varepsilon$ is the energy of the electron, $\varepsilon'=
\omega-\varepsilon$, $n_a$ is the number density of atoms in the medium, 
$\psi(x)$
is the logarithmic derivative of the gamma function, 
$\varrho_c$ is the effective impact parameter, see \cite{L1}. Analytical
expression for $M_{++}^{(1)}$ is found in \cite{BK}.

The influence of the multiple scattering
the process of the coherent photon scattering
illustrated in Figure below
where Im~$M_{++}$ and Re~$M_{++}$ are
given as a function of photon energy $\omega$ in gold
in terms of the amplitude
Im~$M_{++}$ (\ref{3.55}) .
Curves 1 and 3 are Im~$M_{++}$ for $\Delta=$0.4435 $m$ and 0.0387 $m$ 
respectively and curves 2 and 4 are -Re~$M_{++}$ for the same values of
$\Delta$.  The value of -Re~$M$ have the maximum
at $\omega \simeq$220~TeV and 
at $\omega \simeq$80~TeV for $\Delta=$0.4435 $m$ and 0.0387 $m$
respectively.
Similar behaviour has the amplitude $M_{+-}$ \cite{BK}.
For lower value of $\Delta$ 
the interval of contributing
the multiple scattering angles increases, so the effect
manifests itself at a lower energy.
It should be mentioned that value of Im~$M_{+-}$ 
is less than 5
\% of  Im~$M_{++}$ for $\Delta < m$ so that shown amplitudes define
actually the behaviour
of process.The following features are seen.
\begin{description} 
\item The amplitude (and cross section) of the coherent photon scattering  
decreases with photon energy starting from $\omega \sim$1~TeV and approach 
to zero at very high energy.
\item The new property
is the appearance of the real part of the 
coherent photon scattering amplitudes at high energy $\omega$. Similar
property have amplitudes in external field. This will be discussed elsewhere.
\end{description} 

{\bf Acknowledgements}
This work
was supported in part by the Russian Fund of Basic Research under Grant
00-02-18007.

\begin{center}
\epsffile{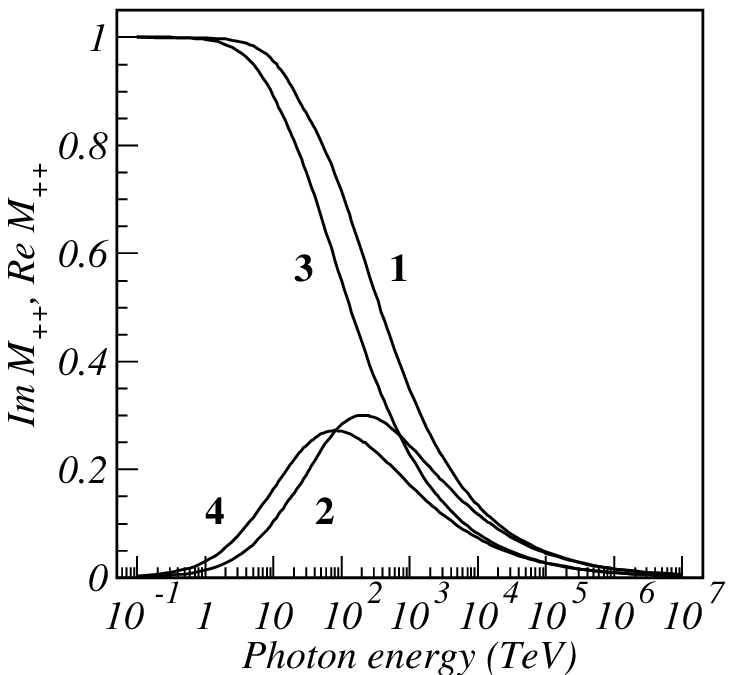}
\end{center}

\end{document}